\documentclass[pra,aps,twocolumn,showpacs,epsfig,pngfig]{revtex4}
\usepackage{color}
\usepackage{graphicx,amssymb,epsfig,epsf,amsmath} 
\begin{document}

\title{Quench dynamics of a Tonks-Girardeau gas in one dimensional anharmonic trap}

\author{Pankaj Kumar Debnath $^{1}$\footnote{e-mail : pankaj\_phys@yahoo.com }, Barnali Chakrabarti $^{2}$\footnote{e-mail : barnali.physics@presiuniv.ac.in},
Mantile Leslie Lekala $^{3}$ }

\affiliation{$^{1}$Swami Vivekananda High School (H.S.); Garshyamnagar II,
P.O.-Shyamnagar, North 24 Parganas-743127, India.}
\affiliation{{$^{2}$ Department of Physics, Presidency University;\\
86/1 College Street, Kolkata 700073, West Bengal, India}}
\affiliation{$^{3}$ Department of Physics, University of South Africa,\\
  P. O. Box 392, Pretoria 0003, South Africa.}


\begin{abstract} 

The quench dynamics of strongly interacting bosons on quartic and sextic traps are studied by solving the time-dependent
many-boson Schr\"odinger equation numerically exactly. The dynamics are addressed by the key measures of one-body density in conjugate
space and information entropy. For both cases, rich many-body dynamics are exhibited and loss of Bose-Fermi oscillation in the Tonks-Girardeau
limit is also attributed.\\

\vspace{.2cm}

Key words: Quench Dynamics; Anharmonic Trap; Tonks-Girardeau Gas; Information Entropy

\end{abstract}

\pacs{ 32.50.+d, 67.85.-d, 89.70.Cf}

\maketitle
\section{Introduction}

The strongly correlated quantum gas is one of the active research areas of nonequilibrium dynamics~\cite{apo,ibl1,ibl2}. In reduced 
dimensions, more intriguing quantum many-body dynamics are expected due to the interplay between the strong interatomic interaction and strong
 correlation. The most seminal observation is the dynamical fermionization. In the fermionization limit the wave function of a Tonks-Girardeau (TG) gas 
of strongly interacting bosons in 1D maps to that of noninteracting fermions in the same potential~\cite{mgi,ehl1,ehl2, dmg}. Dynamical fermionization happens when the momentum distribution approaches asymptotically the Fermi gas distribution during one-dimensional expansion. Whereas an abrupt change in trap frequency exhibits Bose-Fermi oscillation in the momentum distribution~\cite{ami,mri,dmu}. This has also recently been verified experimentally~\cite{jmw}. \\
In the investigation of non-equilibrium dynamics, the temporal responses of interacting bosons are generally explored in the harmonic oscillator potential. However, some references explained the need for anharmonic correction to the harmonic potential~\cite{eha,sgr,jpk}. In the practical situation of experiments, the trap is not purely harmonic, quartic distortion is present~\cite{gql}. It is explicitly shown that the single site of an optical lattice is well represented by a sextic potential, [Fig.1 of Ref.~\cite{sgr} ]. The 1D many-boson systems in anharmonic trap become highly nonintegrable and the corresponding dynamics could be a nontrivial problem. One needs to utilize {\it ab initio} many-body method to solve the time-dependent Schr\"odinger equation. We employ the multiconfigurational time-dependent Hartree method for bosons (MCTDHB)- the exact time-dependent many-body wavefunction is obtained numerically~\cite{oea,oea1}. The system is initially prepared in the harmonic oscillator trap and suddenly quenched to the quartic and sextic confinement. From the computed many-boson wavefunction we calculate the dynamics of one-body density for a wide range of interactions, from weak to TG limit.
The motivation for the work is as follows: 
1) How the quench dynamics in an anharmonic trap is fundamentally different from that of a harmonic trap? Is it possible to observe the Bose-Fermi oscillation in momentum space?  2)  How the many-body features are gradually developed in the dynamical evolution? 3) Does the frequency of breathing oscillation in real space is affected due to a strong correlation? 4) Does the measure of information entropy can describe the dynamical features? \\
Our observations are as follows: a) We find, that the usual Bose-Fermi oscillation is destroyed in the anharmonic trap. b) Clear many-body features are observed in the dynamical evolution. c) We find a strong interplay between trap geometry on the frequency of oscillation. d) Measure of information entropy justifies the dynamical process. 


\section{Model Hamiltonian and set up}
We use dimensionless unit defined by dividing the Hamiltonian by $\frac{\hbar^{2}}{mL^{2}}$, $m$ is the mass of a boson, $L$ is the length scale. The many-body Hamiltonian reads as 

\begin{equation}
    H = \sum^{N}_{i=1} h(x_i) + \sum_{i<j} W(x_i-x_j)
\end{equation}
where $h(x) = -\frac{1}{2} \frac{\partial^2}{\partial x^2}+V(x)$. $V(x)$ is the trapping potential and is assumed as optical lattice $V(x)=V_0 sin^2(\frac{2\pi}{\lambda} x)$. $V_0$ is the depth of the lattice. Restriction to a single well trap, we expand $V(x)$ and cut off the series up to a sextic term in $x$. $V(x)$ reads as 
\begin{equation}
    V(x) = \frac{1}{2} x^2 + \alpha x^{4} +\beta x^6
\end{equation}
Working in the oscillator units where length is defined as $L= 
\sqrt{\frac{{\hbar}}{m\omega}}$ and energy as $\hbar\omega$, it is explicitly shown~\cite{isi} that $\alpha =-0.03$, $\beta= \frac{4 \alpha^2}{5}$ correspond to the trap parameters of Innsbruck experiment~\cite{eha}. $W(x-x^{\prime})$ = $\lambda \delta(x-x^{\prime})$ is the two-body interaction, $\lambda$ is the interaction strength and is determined by the scattering length and the transverse confinement. Throughout our work, we assume repulsive interaction, $\lambda > 0$. \\
In the MCTDHB method, the ansatz for the many-body wave function is the linear combination of time-dependent permanents 
\begin{equation}
\vert \psi(t)\rangle = \sum_{\bar{n}}^{} C_{\bar{n}}(t)\vert \bar{n};t\rangle,
\label{many_body_wf}
\end{equation}
The time dependent permanents $\{ \vert \bar{n}; t \rangle \}$ are obtained by distributing $N$ bosons over $M$ time-dependent orbitals $\{ \phi_i(x,t)\}$. The vector $\vec{n} = (n_1,n_2, \dots ,n_M)$ represents the occupation of the orbitals and $n_1 + n_2 + \dots +n_M = N$ preserves the total number of particles. The expansion coefficients $\{ C_{\bar{n}} (t) \}$ and the orbitals $\{ \phi_i(x,t)\}$, both are time-dependent and are determined by the time-dependent variational principle. 

Compared to the time-independent basis, a given degree of accuracy is reached with a much shorter expansion. From the time-dependent wave function, the reduced one-body density matrix is defined as $\rho^{(1)}(x|x^{\prime};t)$ = $\langle \psi(t) | \hat{\psi}^{\dagger}(x^{\prime}) \hat{\psi}(x) | \psi(t) \rangle$, where $\hat {\psi} (x)$ is the bosonic field operator which annihilates a particle at the position $x$. The diagonal part of one-body density matrix, $\rho(x;t)= \rho^{(1)} ( x|x^{\prime} =x;t)$ is the usual density of the system. Utilizing the Fourier transformation, the one-body density in the momentum space is calculated. This is to be noted that the use of optimized time-dependent orbitals leads to very fast convergence in the simulation compared to the many-body Schr\"odinger equation with time-independent orbitals. It is already established as a very efficient many-body method, numerical convergence to the exact time-dependent solution of the many-body Schr\"odinger equation in different trap geometry is tested~\cite{oea2}. The MCTDHB method and the algorithm have been cast into a software package~\cite{oea3}.\\

We stress here that MCTDHB is much more accurate than
exact diagonalization methods at the same dimensionality
of the considered space. In exact diagonalization, a 
time-independent basis is employed which is built
from the eigenstates of a one-body problem. These states
are not further optimized to take into account the dynamics
and correlations in the considered system, which necessarily
arise due to the presence of interparticle correlation. In this
sense, the Hilbert space and basis used in exact diagonalization
are fixed and not optimized, especially for the treatment of dynamics.

Apart from the one-body density dynamics, another important quantity to characterize the dynamical properties is the many-body Shannon information entropy~\cite{rps,kds}. We calculate many-body information entropy as $S(t)$ = $- \sum_{i} \bar{n}_{i}(t) \large[ ln \bar{n}_{i}(t) \large]$; $\bar{n}_{i} = \frac{n_i}{N}$.
For the Gross Pitaveski mean-field theory, one has $ S(t)$ =0, as there is only one natural occupation $\vec{n}_1=\frac{n_1}{N}=1$ in this case. For multiorbital theories, several occupation numbers can be different from $0$, and the magnitude of  $S$ indicates how well the state could be described by a mean-field approach. The motivation for the calculation of information entropy is to understand the close relation between the time scale of density dynamics and that of many-body information entropy.\\

\begin{figure}
\centering 
\includegraphics[width=0.72 \textwidth, angle=-90]{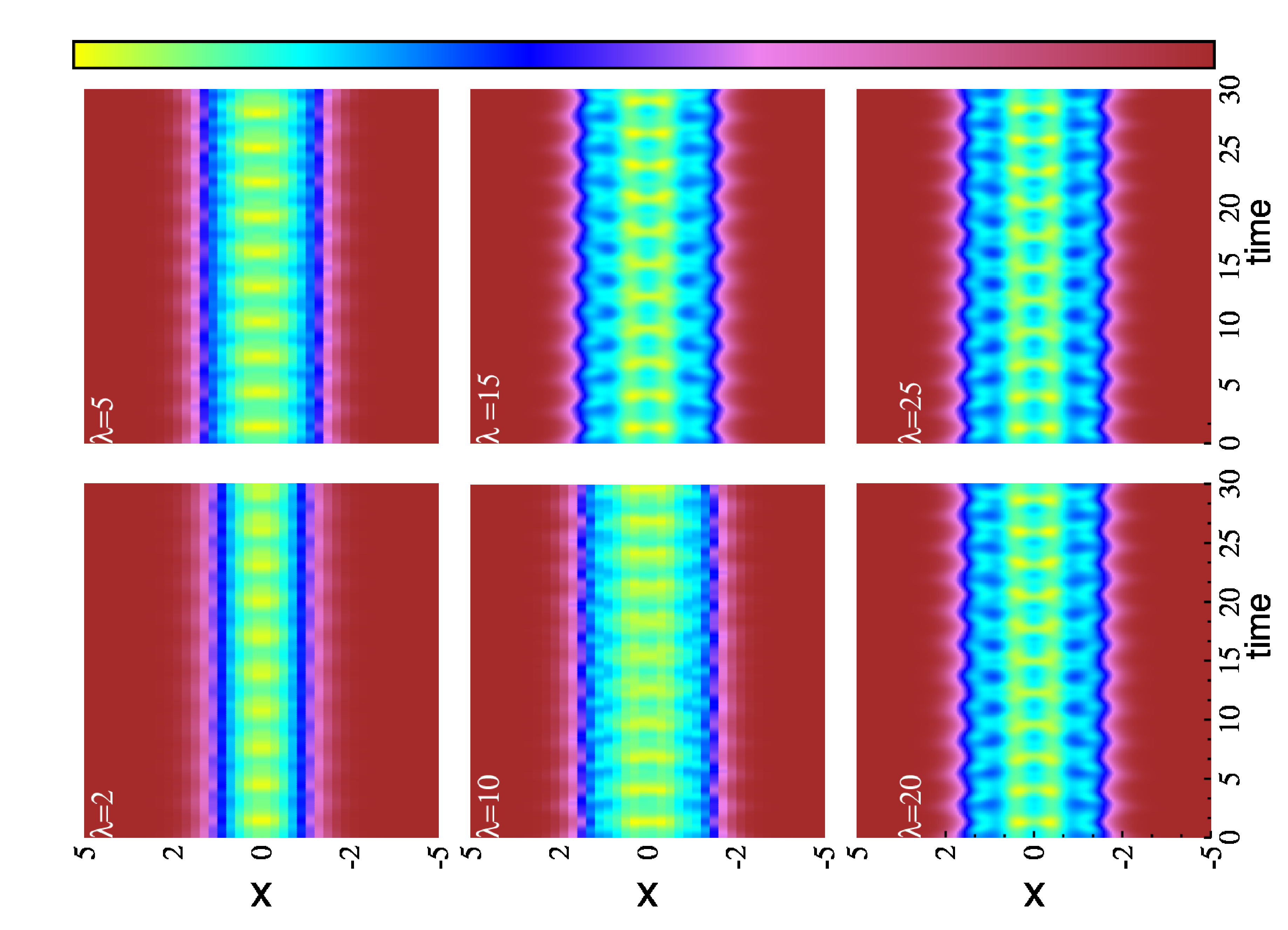}	
\caption{ The figure showcases the captivating time evolution of
the density profile for $N=4$ interacting bosons in the quartic trap with $\alpha=0.02$ and $\beta=0$ and for various interaction strength parameters. Increasing $\lambda$, many-body features are gradually developed and at the TG limit (strongest interaction with $\lambda=25$), the density exhibits exactly four humps which are equal to the number of bosons. Inner humps have maximum modulation and the outer humps have minimum modulation. See the text for further details. The four connected humps exhibit the bounded nature of contact interaction.}
\label{fig1}%
\end{figure}

Thus we investigate the dynamics by the measures of three key quantities: real space or $x$-space density; momentum or $k$-space density and many-body information entropy. To achieve the TG limit, the strongest interaction used in the present calculation is $\lambda =25$, and computation is done with $M=24$ orbitals which restricts the number of bosons to $N=4$ to guarantee the convergence is achieved. We prepare the relaxed state in the harmonic oscillator potential with $\lambda =25$ and the one-body density exactly reproduces the fermionic density distribution in the same trap. The ground state energy of the strongly repulsive bosons also converges to the ground state energy of noninteracting fermions. It guarantees that the choice of $\lambda$ =$25$ leads to the TG limit and can be quenched to an anharmonic trap to understand how the Bose-Fermi oscillation is lost in the
quartic and sextic trap. However, our complete analysis also includes other values of interaction strength parameters. \\

\section{Results}
\subsection{Quench to quartic trap}
The initial state with four interacting bosons is prepared in the ground state of the harmonic oscillator potential $(\alpha=0, \beta=0)$ in Eq.(2) and
quenched to the quartic trap with $\alpha = 0.02$, $\beta=0$. The non-equilibrium dynamics are investigated by the one-body density dynamics
for different choices of interaction strength parameters. The results are presented in Fig.1,  we observe the so-called breathing oscillation.
The dynamics of real space density consist of self-similar cycles with the same frequency for all interaction strength parameters.
Even for the strongest interaction, no damping is observed. We display the results for $\lambda = 2.0, 5.0, 10.0, 15, 20.0$ and $25.0$ to exhibit
how the many-body features are displayed in the dynamics. For $\lambda=2.0$, initially, all four bosons remain in a clustered profile at the minimum
of the oscillator. One-body density is simply a Gaussian profile around the center of the trap. On sudden quench to quartic trap, the height of
the Gaussian peak expands and contracts which is exhibited as a breathing oscillation of constant frequency. The dynamics for much higher
interaction strength $\lambda =5.0$, initially, we find a Gaussian profile at the center of the trap with a small kink in
the middle (in a 2D plot, not shown here). It exhibits the signature of strong interaction without any other many-body features. It exhibits
the same breathing oscillation at the same time scale throughout the evolution. For $\lambda=10.0$, we observe that the density gradually acquires modulations, 
and some signatures of four humps for four interacting bosons start to be generated. However, these four maxima become clear when the interaction
reaches $\lambda =15$. The density modulations are now more pronounced. For much stronger interaction $\lambda=20$ and in the TG
limit ($\lambda=25$) the number of humps saturates to four--the number of bosons in the system. However, the humps are more pronounced at
the center of the trap as exhibited by the yellow dumble-type structure. The density modulation is maximum at the center as the potential is
zero there. However, the yellow dumbles are connected, which clearly demonstrates that the two innermost humps in the density are not isolated.
The two outermost humps--the blue structure on both sides of the yellow co-dumble structure are less pronounced. Due to the larger distance from the
center of the trap, the confining potential is nonzero and the modulation at the outermost peaks is less. The structure is carried on with the same frequency
throughout the dynamics.\\

\begin{figure}
	\centering 
	\includegraphics[width=0.8\textwidth, angle=-90]{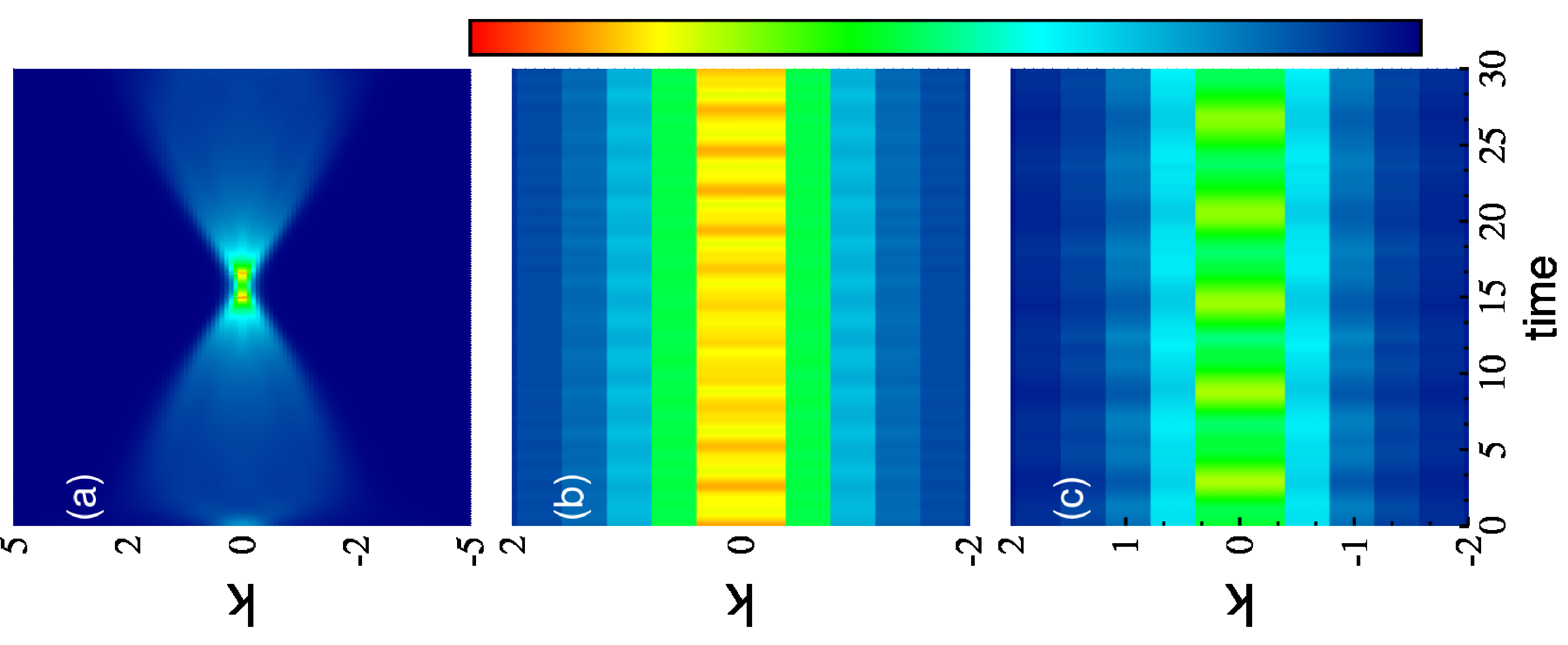}	
	\caption{The momentum-space density dynamics. All results are for $\lambda=25$. The top panel (a) exhibits the Bose-Fermi oscillation when quenched in a harmonic trap of lower frequency. The middle panel (b) presents the density profile for quench to quartic trap and the lower panel (c) exhibits the same for quench to sextic trap. Bose-Fermi oscillation is destroyed in quartic and sextic traps and simple breathing oscillation is exhibited.} 
	\label{fig2}%
\end{figure}
In the well-studied dynamical fermionization, the harmonically confined TG gas displays alternating bosonic and fermionic character in momentum space as shown in Fig.2(a). As before we prepare the initial state of strongly interacting bosons in the TG limit with $\lambda = 25$,  in the same harmonic oscillator potential. Then, we suddenly change the trap frequency to one-tenth of the initial value and measure the density dynamics. We observe a very rich structure in the momentum density, it oscillates between a bosonic-like and fermionic-like momentum distribution. Fig. 2(b) represents the same for quench to quartic trap in TG limit for $\lambda=25$, we observe the Bose-Fermi oscillation is destroyed.
However, the complete eleven cycles are clearly demonstrated in Fig. 2(b).
To assess the dynamical process in the TG limit, we calculate the many-body information entropy $S(t)$ and plot it in Fig. 3 (top panel). We observe an undulated oscillatory nature about the mean value of $S=1.2$, in the same time scale as observed in density dynamics. 

\begin{figure}
	\centering 
	\includegraphics[width=0.84\textwidth, angle=-90]{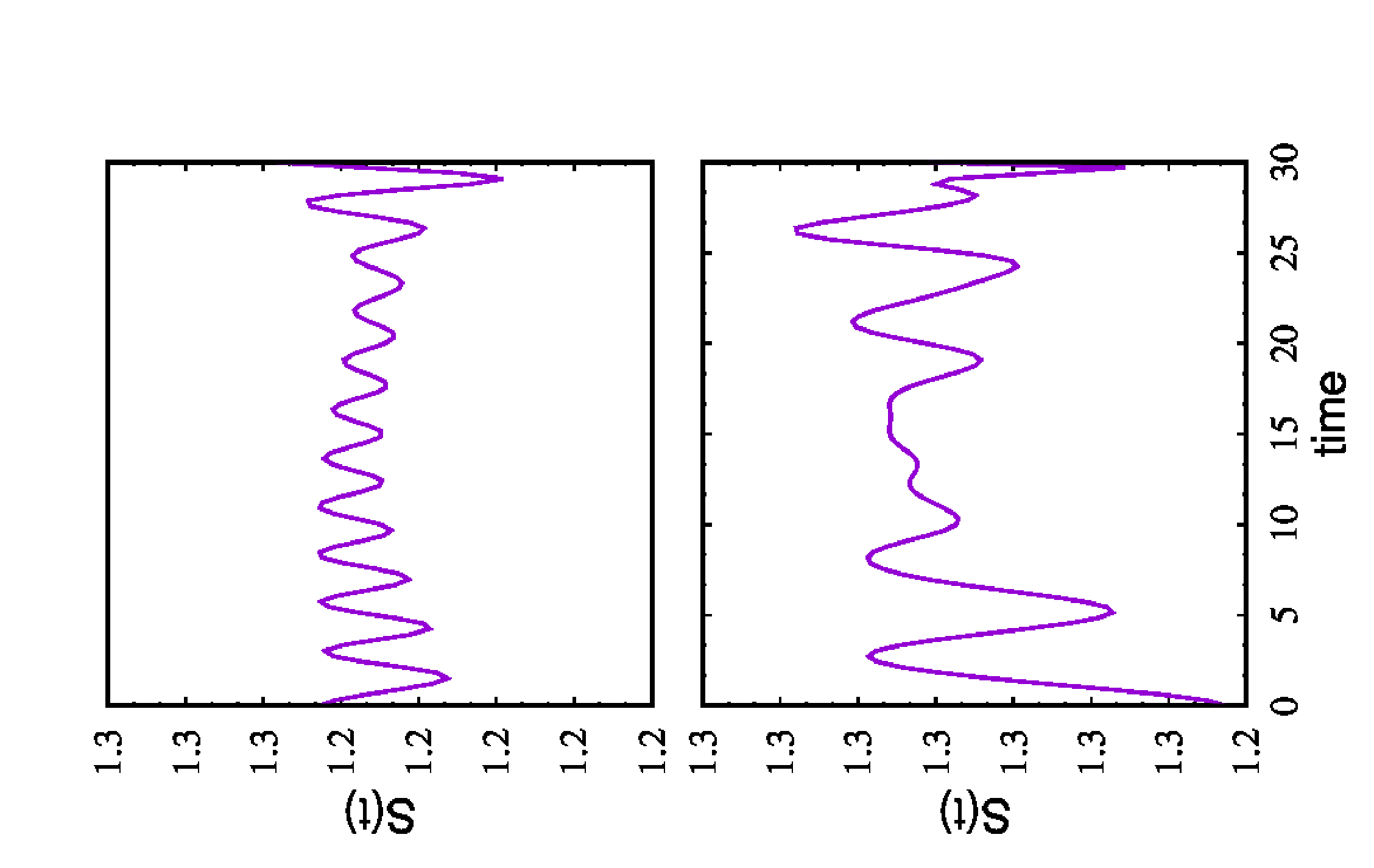}	
	\caption{The Shannon information entropy is plotted as a function of time for $\lambda=25$. The top panel corresponds to the quench to the quartic trap. It exhibits unmodulated oscillation as seen in Fig.1, number of oscillations agrees with that predicted from Fig.1 and Fig.2. The lower panel corresponds to quench to sextic trap. Some irregular nature is observed. However, the number of oscillations exactly matches as predicted in Fig.2 and Fig.4. In both cases the average oscillation value lies in the range of $1.2$ to $1.3$ which agrees with the prediction from random matrix theory. } 
	\label{fig3}%
\end{figure}

We make a comparison between the mean value of numerically evaluated entropy and the prediction of the Gaussian orthogonal ensemble of random matrices (GOE)~\cite{vkb}. For GOE random matrices, $S_{GOE}= ln(M)$, where $M$ is the number of contributing orbitals, not the number of orbitals used in the computation. We keep quite a large number of orbitals $M=24$ to ensure that the dynamics are well converged. However, for $\lambda=25$, as the TG limit is achieved and the strongly interacting bosons behave like noninteracting fermions, the significantly contributing orbitals are just four--exactly equal to the number of bosons. Thus in the TG limit, the many-body state is four-fold fragmented, lowest four natural orbitals contribute, thus $M=4$ and $S_{GOE}=1.38$ which is close to the average value of oscillation in entropy as shown in Fig. 3.  
\begin{figure}
	\centering 
	\includegraphics[width=0.7\textwidth, angle=-90]{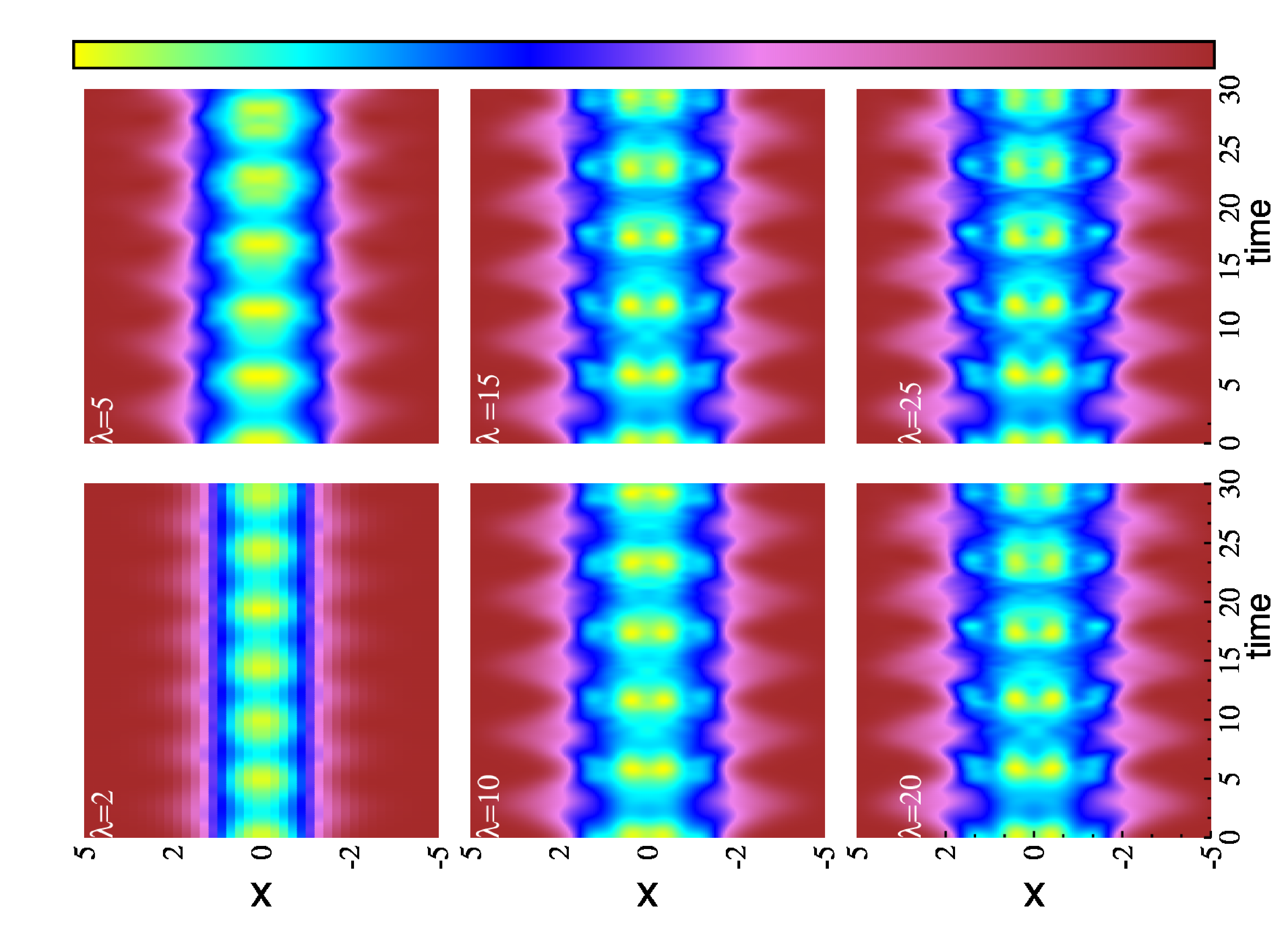}	
	\caption{The figure showcases the captivating time evolution of
the density profile for $N=4$ interacting bosons in the sextic trap with $\alpha=-0.03$ and $\beta=0.00072$ and for various interaction strength parameters. 
For $\lambda=5.0$, the initial Gaussian profile becomes distorted, and many-body features start to develop.
Increasing $\lambda$, many-body features become gradually distinct and at the TG limit (strongest interaction with $\lambda=25$), the density exhibits exactly four humps which are equal to the number of bosons. Inner humps have maximum modulation and the outer humps have minimum modulation. However, the modulated humps are more prominent compared to Fig. 1. See the text for further details. The four connected humps exhibit the bounded nature of contact interaction.} 
	\label{fig4}%
\end{figure}

\subsection{Quench dynamics in sextic trap}
As before we prepare the relaxed state of four interacting bosons in the ground state of harmonic oscillator potential.  Suddenly we quench to a sextic trap as governed by  Eq.(2) with $\alpha=-0.03$ and $\beta=0.00072$. The choices of interaction strength parameters remain the same as those in the previous quench to quartic trap.  The corresponding one-body density dynamics is plotted in Fig. 4 for $\lambda= 2.0, 5.0, 10.0, 15.0, 20.0$ and $25.0$. In comparison with the quench to quartic trap as discussed earlier, our major observations are as follows.\\
a) Frequency of breathing oscillation is increased.
b) Unlike the case of quench to quartic trap, the frequency of oscillation in the sextic trap is affected by the choice of interaction strength. 
c) Many-body features are developed for the lesser value of interaction strength. 
d) For the strongest interaction, the many-body effect is more distinct.\\

Details are as follows. For $\lambda=2.0$, the density remains as Gaussian at the center of the trap. The height of the peak contract and expand
at a much longer scale compared to the quench to quartic trap. For $\lambda=5.0$, we observe some damping effect, and the frequency of breathing oscillation
increases compared to that for $\lambda=2.0$. We also observe that the many-body features start to develop; the presence of two innermost humps is
identified whereas the outermost humps are not clearly seen.\\ 

For $\lambda=10.0$, the innermost and outermost density modulations are clearly visible. There is no further modulation in the oscillatory frequency, no further
damping effect is visible with higher interaction strength. For much higher strength $\lambda =15$, $20$, and $25$, the density modulation becomes more distinct
only, innermost humps are still connected as complete isolation is not possible for contact interaction. The corresponding momentum density as displayed
in Fig. 2(c) for $\lambda=25$ retains the same conclusion made in Fig. 2(b). Bose-Fermi oscillation is destroyed, however, it exhibits a complete number
of oscillations is just five as made in the observation of Fig. 4. The corresponding entropy evolution exhibits the oscillation with an average
value of $1.3$, however, some irregularity is observed compared to regular oscillation as observed for quench to quartic trap. We have checked
the oscillation persists even for quite a long time.\\

\section{Summary and conclusions}
We have presented numerically exact results for many-body dynamics for quenching to quartic and sextic traps. The calculation ranges from weak to very strong interaction in the Tonks-Girardeau limit. We observe that in real space the dynamics exhibit breathing oscillation, however, with an increase in interaction strength many body features are gradually developed. In a quartic trap, the many-body dynamics are unmodulated oscillations and the distinct many-body features are carried on in long-time propagation. Whereas the features are more distinct in a sextic trap, some modulation effect is observed. However, in both cases, there is no signature of Bose-Fermi oscillation even in the long-time dynamics. The entropy measures also exhibit the same physics and the average value of entropy agreed with the prediction of the Gaussian Orthogonal ensemble. \\


\end{document}